\documentclass[10pt,conference]{IEEEtran}

\usepackage{times}
\usepackage{amsmath,latexsym,amsthm,amssymb,amscd,url,enumerate,mathtools}
\usepackage[show]{ed}
\usepackage[all]{xy}
\usepackage{tikz}
\usepackage{multirow}
\usepackage{relsize}

\begin{document} 
\title{Cubical Categories for Higher-Dimensional Parametricity}
\author{\IEEEauthorblockN{Patricia Johann and Kristina Sojakova}
  \\ \IEEEauthorblockA{Appalachian State University, Boone, NC,
    USA\\ email: \{johannp, sojakovak\}@appstate.edu}}
\maketitle

\newcommand{\C}{\mathcal{C}}
\newcommand{\X}{\mathcal{X}}
\newcommand{\Y}{\mathcal{Y}}
\newcommand{\F}{\mathcal{F}}
\newcommand{\G}{\mathcal{G}}
\newcommand{\E}{\mathcal{E}}
\newcommand{\B}{\mathcal{B}}
\newcommand{\U}{\mathbb{U}}
\newcommand{\termobj}{\mathbf{1}}
\newcommand{\fm}{\mathbf{f}}
\newcommand{\dm}{\mathbf{d}}
\newcommand{\nat}{\mathbb{N}}
\newcommand{\two}{\mathbf{2}}
\newcommand{\fst}{\mathsf{fst}}
\newcommand{\snd}{\mathsf{snd}}
\newcommand{\eval}{\mathsf{eval}}
\newcommand{\cat}{\mathsf{Cat}}
\newcommand{\rel}{\mathsf{Rel}}
\newcommand{\set}{\mathsf{Set}}
\newcommand{\Ob}{\mathbf{Ob}}
\newcommand{\Mor}{\mathbf{Mor}}
\newcommand{\ctx}{\mathsf{Ctx}}
\newcommand{\Set}{\mathsf{Set}}
\newcommand{\Prop}{\mathsf{Prop}}
\newcommand{\sem}[1]{\ensuremath{[\![ #1 ]\!]}}
\newcommand{\grph}[1]{\ensuremath{\langle #1 \rangle}}
\newcommand{\cart}[1]{\ensuremath{#1^{\S}}}
\newcommand{\opcart}[1]{\ensuremath{#1_{\S}}}
\newcommand{\foldd}[3][]{\ensuremath{\mathit{fold}_{#1}[#2, #3]}}
\newcommand{\inn}[1][]{\ensuremath{\mathit{in}_{#1}}}
\newcommand{\id}{\mathsf{id}}
\newcommand{\Eq}{\mathsf{Eq}}
\newcommand{\eq}{\mathsf{eq}}
\newcommand{\expo}[2]{\ensuremath{#1 \Rightarrow #2}}
\newcommand{\internal}[1]{\ensuremath{\underline{#1}}}
\newcommand{\Id}{\mathsf{Id}}

\newtheorem{theorem}{Theorem}						
\newtheorem{definition}[theorem]{Definition}
\newtheorem{proposition}[theorem]{Proposition}
\newtheorem{lemma}[theorem]{Lemma}
\newtheorem{remark}[theorem]{Remark}
\newtheorem{example}[theorem]{Example}
\newtheorem{notation}[theorem]{Notation}
\newtheorem{corollary}[theorem]{Corollary}


\begin{abstract}
Reynolds' theory of {\em relational parametricity} formalizes
parametric polymorphism for System F, thus capturing the idea that
polymorphically typed System F programs always map related inputs to
related results. This paper shows that Reynolds' theory can be seen as
the instantiation at dimension $1$ of a theory of relational
parametricity for System F that holds at all higher dimensions,
including infinite dimension. This theory is formulated in terms of
the new notion of a {\em $p$-dimensional cubical category}, which we
use to define a {\em $p$-dimensional parametric model} of System F for
any $p \in \nat \cup \{\infty\}$. We show that every $p$-dimensional
parametric model of System F yields a split $\lambda 2$-fibration in
which types are interpreted as face map- and degeneracy-preserving
cubical functors and terms are interpreted as face-map and
degeneracy-preserving cubical natural transformations.  We demonstrate
that our theory is ``good'' by showing that the PER model of
Bainbridge {\em et al.} is derivable as another $1$-dimensional
instance, and that all instances at all dimensions derive
higher-dimensional analogues of expected results for parametric
models, such as a Graph Lemma and the existence of initial algebras
and final coalgebras. Finally, our technical development resolves a
number of significant technical issues arising in Ghani {\em et al.}'s
recent bifibrational treatment of relational parametricity, which
allows us to clarify their approach and strengthen their main
result. Once clarified, their bifibrational framework, too, can be
seen as a $1$-dimensional instance of our theory.
\end{abstract}

\section{Introduction}

Strachey~\cite{str00} distinguished between {\em ad hoc} and
parametric polymorphic functions in programming languages, defining a
polymorphic program to be {\em parametric} if it applies the same
type-uniform algorithm at each of its type instantiations.
Reynolds~\cite{param_reynolds} introduced the notion of {\em
  relational parametricity} to model the extensional behavior of
parametric programs in System F~\cite{systemf}, the formal calculus at
the core of polymorphic functional languages.  Relationally parametric
models capture a key feature of parametric programs, namely that they
preserve all relations between instantiated types. In other words, in
relationally parametric models, parametric polymorphic functions
always map related arguments to related results.

Implicit in Reynolds' original formulation of relational
parametricity~\cite{param_reynolds} is that a model of System F is
{\em relationally parametric} if equality in the model is induced by a
logical relation. A {\em logical relation} assigns to each type of a
language not only a basic interpretation as, say, a set or a domain,
but simultaneously an interpretation as a relation on that set or
domain as well.  Logical relations are defined by induction on the
language's type structure, and are constructed in such a way that the
relational actions interpreting its type formers propagate relatedness
up its type hierarchy. For each logical relation for a language,
a {\em parametricity theorem} can then be proved.
Such a theorem states that (the basic interpretation of) each of the
languages' programs is related to itself by the relational
interpretation, via the associated logical relation, of that program's
type. When instantiated judiciously, this seemingly simple result can
be used to prove, {\em inter alia}, invariance of polymorphic
functions under changes of data representation~\cite{adr09,dnb12},
equivalences of programs~\cite{hd11}, and so-called ``free theorems''
via which properties of programs can be inferred solely from their
types~\cite{wad87}.

The recent bifibrational treatment of relational parametricity
in~\cite{param_johann} has put forth a more abstract notion of a
parametric model of polymorphism. In this treatment every type is
still given a interpretation in a sufficiently structured base
category, together with a relational interpretation in a category of
(now abstractly formulated) relations over that base category, but the
two interpretations are defined simultaneously and are required to be
connected via a sufficiently structured bifibration. The express aim
of~\cite{param_johann} is to provide a very general framework for
relational parametricity that is directly instantiable not only to
recover well-known relationally parametric models --- such as
Reynolds' original model\footnote{Since there are no set-theoretic
  models of System F, by the phrase ``Reynolds' original model'' we
  will mean the version of his model that is internal to the Calculus
  of Inductive Constructions with Impredicative Set (as indicated
  in~\cite{param_johann}).} and the PER model of Bainbridge {\em et
  al.}  --- but also to deliver entirely new models of relational
parametricity for System F.

Unfortunately, however, models of relational parametricity often
require more careful notions of functor and natural transformation
than just the standard categorical ones used in the bifibrational
framework of~\cite{param_johann}. For example, functors and natural
transformations must be internal to the category of types and terms in
the Calculus of Inductive Constructions with impredicative $\set$ to
recover Reynolds' original model, and they must be internal to the
category of $\omega$-sets to recover Bainbridge {\em et al.}'s PER
model. As a result, neither of these models is a true instance of the
bifibrational framework of~\cite{param_johann}. Said differently, the
bifibrational framework of~\cite{param_johann} is not actually an
extension of Reynolds' theory of relational parametricity as
claimed. In fact, showing that Reynolds' original model is parametric
in the sense specified by the bifibrational framework requires a
complete redevelopment of the framework internally to the Calculus of
Inductive Constructions with impredicative $\set$, and showing that
the PER model is parametric requires a redevelopment internal to the
category of $\omega$-sets. The need to redevelop the entire framework
of~\cite{param_johann} internal to a different category for each
relationally parametric model of interest makes the bifibrational
framework more of a ``blueprint'' for constructing parametric models
than a general theory that actually includes known models properly among its
instances. The fact that such redevelopments must also be carried out
on an {\em ad hoc} basis, without any generally-applicable guidance,
only emphasizes the need for a truly instantiable theory of relational
parametricity.

But even if uniform guidance for instantiating the framework
of~\cite{param_johann} {\em were} to be given, the framework itself
would still be problematic. Unless fibred functors are required to
preserve equality on the nose, neither composition nor substitution in
(what is intended to be) the base category of the $\lambda
2$-fibration constructed in the main theorem of~\cite{param_johann} can 
be defined in any standard way. But equality in~\cite{param_johann} is only defined
--- and therefore can only be preserved --- up to isomorphism. And
even if the original bifibration from which the $\lambda 2$-fibration 
in~\cite{param_johann} is constructed ($U$ in the terminology
there) were assumed to be split, so that the equality
functor were defined uniquely rather than only up to isomorphism,
Reynolds' original model 
still would not be an
instance of the bifibrational framework given there: in that case,
neither products nor exponentials would preserve equality on the nose,
as would be needed to properly interpret arrow types.  In the absence
of any alternative definitions or discussion of the exact sense in
which fibred functors are required to preserve equality, we can only
conclude that the standard definitions are the intended ones.  As a
result, we regard the entire $\lambda 2$-fibration as being
ill-defined, and the beautiful ideas explored in~\cite{param_johann}
as being in need of careful technical reconsideration.

This paper provides precisely such a reconsideration, as well as a
significant extension. We remedy both of the aforementioned
difficulties by developing a unifying approach to relational
parametricity that turns the bifibrational ``blueprint" for
constructing parametric models for System F given by the framework
of~\cite{param_johann} into a single theory whose instantiation
actually delivers such models. Our theory combines two key technical
ingredients to produce $\lambda 2$-fibrations that not only are
actually well-defined, but do really model relational
parametricity. First, we ensure that the categories necessary to our
constructions are well-defined by parameterizing our theory over a
class of ``good'' natural isomorphisms, and requiring that fibred
functors preserve equality only up to these isomorphisms and
(essentially) that fibred functors transformations preserve them.
Secondly, we ensure that well-known models of relational parametricity
for System F are properly instances of our theory by parameterizing it
over a suitably structured ambient category and working internally to
that category.  But working internally to an appropriate ambient
category is more than just a technical device ensuring that all of our
constructions are well-defined and that well-known parametric models
for System F are instances of our theory. It is also {\em precisely}
the mechanism by which we restrict the possible interpretations of
types and terms in our $\lambda 2$-fibrations sufficiently to exclude
{\em ad hoc} polymorphism. This is illustrated concretely in
Example~\ref{ex:per} below.

In addition to remedying all known problems with the bifibrational
framework of~\cite{param_johann}, the theory we develop in this paper
also naturally opens the way to a theory of relational parametricity
at higher dimensions. Indeed, our theory deploys the two key
ingredients identified above not solely in a bifibrational setting
similar to that in~\cite{param_johann}, but also in combination with
ideas inspired by the theory of cubical
sets~\cite{cubical_coquand,cubical_kan,cubical_harper}.  In this way,
it delivers $\lambda 2$-fibrations that model more than just the
single ``level'', or ``dimension'', of relational parametricity
originally identified by Reynolds and considered
in~\cite{param_johann}. To enforce relational parametricity at higher
dimensions we introduce the new notion of a {\em $p$-dimensional
  cubical category}, in terms of which we define the equally new
notion of a {\em $p$-dimensional parametric model} for System F. Here,
the dimension $p$ can be any natural number or $\infty$. 
Intuitively, cubical categories generalize cubical sets in the obvious way,
by considering the codomain $\cat$ instead of $\set$. Technically, the codomain
will be $\cat(\C)$, the category of categories internal to some sufficiently
structured ambient category $\C$.
Cubical categories have
(essentially) the same algebraic structure as cubical sets, except
that the morphisms in their domain category are restricted to just
those generated by face maps and degeneracies, {\em i.e.}, to just
those have natural interpretations as operations on relations. This
ensures that morphisms are restricted to those that, intuitively, have
interpretations as operations on relations.

Our main technical result (Theorem~\ref{thm:split-lambda2-fib}) shows
that every $p$-dimensional parametric model of System F gives rise to
a split $\lambda 2$-fibration. When combined with a suitable variant
of Seely's result that every split $\lambda 2$-fibration gives rise to
a sound model of System F (Theorem~\ref{prop:model}), this allows us
to prove that every $p$-dimensional parametric model of System F gives
a sound model of that calculus in which types are interpreted as
$p$-dimensional face map- and degeneracy-preserving cubical functors
and terms are interpreted as $p$-dimensional face map-preserving and
degeneracy-preserving cubical natural transformations. This, our main
result, appears as Theorem~\ref{thm:cubical-model} below. It
strengthens the analogous result in~\cite{param_johann}, which states
that natural transformations interpreting terms must be face
map-preserving when $p = 1$, but does not observe that they can also
be proved to be degeneracy-preserving even in the $1$-dimensional
setting. Because they interpret System F terms as face map- and
degeneracy-preserving cubical natural transformations, thereby
ensuring that these terms cannot exhibit {\em ad hoc} polymorphic
behavior, we contend that $p$-dimensional parametric models of System
F are deserving of their name. That both Reynolds' model and that of
Bainbridge {\em et al.} are both $p$-dimensional parametric models for
System F when $p = 1$ further shows our definition is both sensible
and good. Additional evidence is provided in
Section~\ref{sec:consequences}, where it is shown that all
$p$-dimensional parametric models of System F validate
higher-dimensional analogues of the ``litmus test'' properties for
``good'' parametric models. That is, they validate a
higher-dimensional Identity Extension Lemma, a higher-dimensional
Graph Lemma, and the existence of initial algebras and final
coalgebras for face map- and degeneracy-preserving cubical functors.

Instantiations of our theory for specific choices for $p$ may already
be of particular interest. When $p = 2$, our notion of a
$p$-dimensional parametric model of System F formalizes a notion of
proof-relevant relational parametricity that properly
generalizes Reynolds' original theory.
When $p = \infty$, we get a notion of infinite-dimensional relational
parametricity for System F that may provide a useful perspective on
the homotopy-canonicity conjecture for homotopy type theory, since
proof of this conjecture involves constructing an infinitely
parametric model of Martin-L\"of type theory. Investigation of this
matter is, however, beyond the scope of the present paper.

\section{Fibrational Preliminaries}\label{sec:toolbox}

We give a brief introduction to fibrations, mainly to settle
notation. More details can be found in, e.g.,~\cite{jac99}.

\begin{definition}
Let $U:\mathcal{E}\rightarrow \mathcal{B}$ be a functor.  A morphism
$g:Q\rightarrow P$ in $\mathcal{E}$ is \emph{cartesian} over
$f:X\rightarrow Y$ in $\mathcal{B}$ if $Ug=f$ and, for every
$g':Q'\rightarrow P$ in $\mathcal{E}$ with $Ug' = f \circ v$ for some
$v:UQ'\rightarrow X$, there exists a unique $h:Q'\rightarrow Q$ with
$Uh=v$ and $g' = g \circ h$.  A morphism $g:P\rightarrow Q$ in
$\mathcal{E}$ is \emph{opcartesian} over $f:X\rightarrow Y$ in
$\mathcal{B}$ if $Ug=f$ and, for every $g':P\rightarrow Q'$ in
$\mathcal{E}$ with $Ug'=v \circ f$ for some $v:Y\rightarrow UQ'$,
there exists a unique $h:Q\rightarrow Q'$ with $Uh=v$ and $g' = h
\circ g$.
\end{definition}

\noindent We write $\cart{f}_{P}$ for the cartesian morphism over $f$
with codomain $P$, and $\opcart{f}^{P}$ for the opcartesian morphism
over $f$ with domain $P$. Such morphisms are unique up to
isomorphism. If $P$ is an object of $\mathcal{E}$ then we write
$f^{\ast}P$ for the domain of $\cart{f}_{P}$ and $\Sigma_{f}P$ for the
codomain of $\opcart{f}^{P}$. We omit $P$ from these notations when it
is either unimportant or clear from context.

\begin{definition}
A functor $U:\mathcal{E}\rightarrow \mathcal{B}$ is a \emph{fibration}
if for every object $P$ of $\mathcal{E}$ and every morphism
$f:X\rightarrow UP$ of $\mathcal{B}$, there is a cartesian morphism
$\cart{f}_{P}:Q\rightarrow P$ in $\mathcal{E}$ over f. Similarly, $U$
is an \emph{opfibration} if for every object $P$ of $\mathcal{E}$ and
every morphism $f:UP\rightarrow Y$ of $\mathcal{B}$, there is an
opcartesian morphism $\opcart{f}^{P}:P\rightarrow Q$ in $\mathcal{E}$
over $f$. A functor $U$ is a \emph{bifibration} if it is both a
fibration and an opfibration.
\end{definition}

\noindent
If $U:\mathcal{E}\rightarrow \mathcal{B}$ is a fibration, opfibration,
or bifibration, then $\mathcal{E}$ is its \emph{total category} and
$\mathcal{B}$ is its \emph{base category}. An object $P$ in
$\mathcal{E}$ is \emph{over} its image $UP$ and similarly for
morphisms. A morphism is {\em vertical} if it is over $id$.  We write
$\mathcal{E}_{X}$ for the \emph{fiber over} an object $X$ in
$\mathcal{B}$, i.e., the subcategory of $\mathcal{E}$ of objects over
$X$ and morphisms over $\mathit{id}_X$. For $f:X\rightarrow Y$ in
$\mathcal{B}$, the function mapping each object $P$ of $\mathcal{E}$
to $f^*P$ extends to a functor $f^*: \mathcal{E}_{Y} \rightarrow
\mathcal{E}_{X}$ mapping each morphism $k:P\rightarrow P'$ in
$\mathcal{E}_{Y}$ to the morphism $f^*k$ with $k \cart{f}_{P} =
\cart{f}_{P'} f^*k$. The universal property of $\cart{f}_{P'}$ ensures
the existence and uniqueness of $f^*k$.  We call $f^*$ the
\emph{reindexing functor along $f$}. A similar situation holds for
opfibrations; the functor $\Sigma_{f}:\mathcal{E}_{X}\rightarrow
\mathcal{E}_{Y}$ extending the function mapping each object $P$ of
$\mathcal{E}$ to $\Sigma_{f}P$ is the \emph{opreindexing functor along
  $f$}.


In this paper we will construct a certain kind of fibration, called a
{\em $\lambda 2$-fibration}~\cite{see87}, that models
higher-dimensional parametricity for System F. Fibrations will also be
essential to defining a higher-dimensional graph functor, and
bifibrations will be crucial to formulating an alternative
characterization of the graph functor that allows us to prove both a
higher-dimensional Graph Lemma, and the existence of initial algebras
and final coalgebras of face map- and degeneracy-preserving cubical
functors.

\section{Cubical Categories}\label{sec:cubical-categories}

Cubical categories, functors, and natural transformations are the key
structures from which we will construct our models of higher-dimensional parametricity. To
define them we start with the following preliminary category.

\begin{definition}
The category $\square$ is given as follows:
\begin{itemize}
\item the objects are (finite) sets of natural numbers of the form $\{
  0, \ldots, l - 1\}$ for $l \in \nat$
\item the morphisms from $l_1$ to $l_2$ are functions from $l_1$ to $l_2 +
  \two$, where $\two$ is the two-element set $\{\top,\bot\}$
\item the identity morphism on $l$ is (induced by) the identity
  function on $l$, {\em i.e.}, is the inclusion map $l \hookrightarrow
  l + \two$
\item the composition of two morphisms $f : l_1 \to l_2$ and $g : l_2
  \to l_3$ is the function $g \circ f : l_1 \to l_3 + \two$ defined by
\begin{align*}
(g \circ f)(i) =
\begin{cases} 
\star & \mbox{if } f(i) = \star, \star \in \two \\ 
g(j) & \mbox{if } f(i) = j, j \in \nat
\end{cases}
\end{align*}
\end{itemize}
\end{definition}

We will henceforth denote a set $\{ 0, \ldots, l - 1\}$ of natural
numbers by $l$. (Set-theoretically, these are identical.) We call any
such set, \emph{i.e.}, any object of $\square$, a \emph{level}. The
category $\square$ can also be described as the Kleisli category for
the error monad with two distinct error values. The last bullet point
then defines composition of morphisms in $\square$ to be normal
composition of functions, except that errors are propagated.

The category $\square$ contains {\em all} functions from $l_1$ to $l_2
+ \two$ for all $l_1$ and $l_2$. But to model parametricity, we will
want to restrict the set of morphisms to those that, intuitively, have
interpretations as operations on relations. For this reason, only the
face maps and degeneracies defined below, and their compositions, are
used to construct our cubical categories. The interpretations of the
face maps and degeneracies in the specific setting
of~\cite{param_johann} are given in Example~\ref{ex:bifib-param}.

For any $\star \in \two$ and $l, k \in \nat$ with $k \leq l$, we
define the function $\fm_\star(l,k) : l+1 \to l$ by
\begin{align*}
\fm_\star(l,k)(i) =
\begin{cases}
i & \mbox{if } i < k \\ 
\star & \mbox{if } i = k \\ i-1 & \mbox{if } i > k
\end{cases}
\end{align*}
Such a function is called a \emph{face map}. The terminology comes
from regarding each natural number $l$ as defining an $l$-dimensional
cube. A face map $\fm_\star(l,k)$ then can be thought of as projecting
an $(l+1)$-cube onto either the ``top'' or the ``bottom''
$l$-dimensional cube in dimension $k$, according as $\star$ is $\top$
or $\bot$. Similarly, for any $l, k \in \nat$ with $k \leq l$, we
define the function $\dm(l,k) : l \to l+1$ by
\begin{align*}
\dm(l,k)(i) =
\begin{cases}
i & \mbox{if } i < k \\ 
i+1 & \mbox{if } i \geq k
\end{cases}
\end{align*}
Such a function is called a \emph{degeneracy}. A degeneracy $\dm(l,k)$
can be thought of as constructing an $(l+1)$-dimensional cube from an
$l$-dimensional one by replicating it along dimension~$k$.

We also have the following set of categories $\square_p$:

\begin{definition}
Let $p \in \nat \cup \{\infty\}$. The category $\square_p$ is the
subcategory of $\square$ generated by the following data:
\begin{itemize}
\item levels $l \leq p$
\item face maps $\fm_\star(l,k)$ for $l < p$
\item degeneracies $\dm(l,k)$ for $l < p$
\end{itemize}
Alternatively, we can define $\square_p$ as the free category
generated by the the data above and the following relations:
\begin{itemize}
\item $\fm_\star(l,k) \circ \dm(l,k) = \mathsf{1}_l$ for $l < p$
\item $\fm_\star(l,j) \circ \dm(l,k) = \dm(l-1,k) \circ
  \fm_\star(l-1,j-1)$ for $k < j$ and $l < p$ 
\item $\fm_\star(l,j) \circ \dm(l,k) = \dm(l-1,k-1) \circ
  \fm_\star(l-1,j)$ for $j < k$ and $l < p$ 
\item $\dm(l+1,j) \circ \dm(l,k) = \dm(l+1,k) \circ \dm(l,j-1)$ for $k
  < j$ and $l+1 < p$ 
\item $\fm_{\star_2}(l,j) \circ \fm_{\star_1}(l+1,k) =
  \fm_{\star_1}(l,k) \circ \fm_{\star_2}(l+1,j+1)$ for $k \leq j$ and
  $l+1 < p$ 
\end{itemize}
\end{definition}

The alternative characterization coincides exactly with Crans'
combinatory treatment of cubes. As proved in~\cite{pasting_crans}, any
morphism in $\square_p$ can be factored as a composition of face maps
followed by a composition of degeneracies. Moreover, the second
arguments to the degeneracies in these compositions are
non-increasing, and the second arguments to the face maps are strictly
decreasing, when read in composition order, {\em i.e.}, from right to
left. Such a factorization gives a representation of each morphism as
a surjection followed by an inclusion, as well as a canonical form for
each morphism in $\square_p$.

The categories $\square_p$ will serve as the domains of our cubical
categories. As such, they are our analogues of the category of names
and substitutions, which forms the common domain of all cubical sets
in~\cite{cubical_coquand}. The differences between our categories
$\square_p$ and the category of names and substitutions are that
$\square_p$ does not include the ``exchange morphisms''
of~\cite{cubical_coquand} (and~\cite{cubical_harper}), and that
membership in $\square_p$ does not explicitly require morphisms to be
injective. That all morphisms in each $\square_p$ are, in fact,
injective follows from the injectivity of $\square_p$'s generators.

In the remainder of this paper we will always work internally with
respect to a finitely complete locally small ambient category $\C$,
{\em i.e.}, a category $\C$ with pullbacks and a terminal object
$\termobj_\C$.\footnote{The requirement that $\C$ has {\em all}
  pullbacks is actually stronger than necessary. In fact, we need only
  require $\C$ to have all ``composable'' pullbacks.} Instantiating
$\C$ appropriately will impose conditions on functors and natural
transformations that allow us to produce a theory of
higher-dimensional parametricity that subsumes well-known models as
($1$-dimensional) relational parametricity as instances of our
framework. For example, Reynolds' original model is an instance when
$\C$ is taken to be the category of types and terms in the Calculus of
Inductive Constructions with impredicative $\set\,$\footnote{The quite
  surprising fact that this theory has not yet actually been proved
  consistent is, however, worth noting.}, and the PER model of
Bainbridge {\em et al.} is an instance when $\C$ is the category of
$\omega$-sets; see Examples~\ref{ex:reynolds} and~\ref{ex:per} below
for details.

Let $\cat(\C)$ be the category of categories internal to an
ambient category $\C$.

\begin{definition}\label{def:cubical-category}
A \emph{($p$-dimensional) cubical category} is a functor $\X :
\square_p \to \cat(\C)$.
\end{definition}
\noindent
To ease the notational and conceptual burden, we will henceforth
regard a ($p$-dimensional) cubical category as a functor $\X :
\square_p \to \cat$, and similarly identify internal and external
constructions when convenient. Under this identification, a
$p$-dimensional cubical category $\X$ becomes the category-level
equivalent of a covariant presheaf on $\square_p$, and gives us a
(small) category $\X(l)$ for each level $l \in \square_p$. The
category $\X(0)$ can be thought of as an abstract category of
``0-relations", or ``types"; $\X(1)$ can be thought of as the category
of ``$1$-relations", or ordinary relations, on types; $\X(2)$ can be
thought of as the category of ``$2$-relations"; and so on. Each face
map $f : l+1 \to l$, $\X(f)$ is thus a functor projecting
an $l$-relation out of a given $(l+1)$-relation, and each
degeneracy $d : l \to l+1$, $\X(d)$ is a functor that
replicates a given $l$-relation to obtain an $(l+1)$-relation.

\begin{example}\label{ex:bifib-param}
In the setting of \cite{param_johann}, a relations fibration $\rel(U)
: \rel(\mathcal{E}) \to \mathcal{B} \times \mathcal{B}$ induces a
$1$-dimensional cubical category, with the action on objects given by $0
\mapsto \mathcal{B}$ and $1 \mapsto \rel(\mathcal{E})$. The action on
morphisms is induced by mapping the two face maps $\fm_\top(0,0)$ and
$\fm_\bot(0,0)$ to the functors $\fst \mathop{\circ} \rel(U)$ and
$\snd \mathop{\circ} \rel(U)$ respectively, and mapping the degeneracy
$\dm(0,0)$ to the equality functor $\mathsf{Eq}$ on $\mathcal{B}$.  
\end{example}

If $\X$ is a cubical category we define the discrete cubical category
$|\X|$, and the product cubical category $\X^n$ for $n \in \nat$, via
the usual constructions for functors. Each construction on cubical
categories actually requires an analogous construction on
$\cat(\C)$, but these are precisely as expected.
For example, for cubical categories $\X, \Y : \square_p \to
\cat(\C)$, the cubical category $\X \times \Y : \square_p \to
\cat(\C)$ is defined by $(\X \times \Y)(l) = \X(l) \times
\Y(l)$ for all $l \leq p$.  Here, the product on the left-hand side is
a product of functors, and the product on the right-hand side is a
product of internal categories. The latter exists because $\C$ is
finitely complete by assumption.

\begin{definition}\label{def:cubical-functor}
Let $\X$ and $\Y$ be ($p$-dimensional) cubical categories. A
\emph{($p$-dimensional) cubical functor} $\F$ from $\X$ to $\Y$ is a
set of functors $\{\F(l) : \X(l) \to Y(l) \; | \; l \leq p \}$. A
cubical functor $\F$ is \emph{face map-preserving} if the following
diagram commutes for every face map $h : l_1 \to l_2$ in $\square_p$:
\begin{center}
\scalebox{0.75}{
\begin{tikzpicture}
\node (N0) at (0,2) {$\X(l_1)$};
\node (N1) at (0,0) {$\X(l_2)$};
\node (N2) at (4,2) {$\Y(l_1)$};
\node (N3) at (4,0) {$\Y(l_2)$};
\draw[->] (N0) -- node[left]{$\X(h)$} (N1);
\draw[->] (N0) -- node[above]{$\F(l_1)$} (N2);
\draw[->] (N1) -- node[below]{$\F(l_2)$} (N3);
\draw[->] (N2) -- node[right]{$\Y(h)$} (N3);
\end{tikzpicture}}
\end{center}
A cubical functor $\F$ is \emph{degeneracy-preserving} if the diagram
above commutes up to a chosen natural isomorphism $\varepsilon_\F(h)$
for each degeneracy $h : l_1 \to l_2$ in $\square_p$.
\end{definition}

In the setting of \cite{param_johann}, a fibred functor from
$|\rel(U)|^n$ to $\rel(U)$ is precisely a face map-preserving cubical
functor from $|\rel(U)|^n$ to $\rel(U)$ (presented as cubical
categories). Similarly, an equality-preserving fibred functor from
$|\rel(U)|^n$ to $\rel(U)$ is a face map-preserving cubical functor
that also preserves degeneracies up to ``good'' natural isomorphisms
(for a suitable notion of ``good''; see Section~\ref{sec:model}
below).

\begin{definition}\label{def:cubical-nat-trans}
Let $\F$ and $\G$ be ($p$-dimensional) cubical functors from $\X$ to
$\Y$. A \emph{($p$-dimensional) cubical natural transformation} $\eta$
from $\F$ to $\G$ is a set of natural transformations $\{\eta(l) :
\F(l) \to G(l) \; | \; l \leq p \}$. A cubical natural transformation
$\eta$ is \emph{face map-preserving} if $\F$ and $\G$ are face
map-preserving and, for each face map $h : l_1 \to l_2$ in
$\square_p$, the following equality holds for every object $X$ of
$\X(l_1)$:
\[ \Y(h) \; \big(\eta(l_1) \; X\big) = \eta(l_2) \; \big(\X(h) \; X\big) \]
A cubical natural transformation $\eta$ is
\emph{degeneracy-preserving} if $\F$ and $\G$ are
degeneracy-preserving and, for each degeneracy $h : l_1 \to l_2$ in
$\square_p$, the following diagram commutes for every object $X$ of
$\X(l_1)$:

\begin{center}
\scalebox{0.75}{
\begin{tikzpicture}
\node (N0) at (0,2.5) {$\Y(h)\big(\F(l_1) \; X\big)$};
\node (N1) at (0,0) {$\Y(h)\big(\G(l_1) \; X\big)$};
\node (N2) at (6,2.5) {$\F(l_2)\big(\X(h) \; X\big)$};
\node (N3) at (6,0) {$\G(l_2)\big(\X(h) \; X\big)$};
\draw[->] (N0) -- node[left]{$\Y(h) \; \big(\eta(l_1) \; X\big)$} (N1);
\draw[->] (N0) -- node[above]{$\varepsilon_\F(h) \; X$} (N2);
\draw[->] (N1) -- node[below]{$\varepsilon_\G(h) \; X$} (N3);
\draw[->] (N2) -- node[right]{$\eta(l_2) \; \big(\X(h) \; X\big)$} (N3);
\end{tikzpicture}}
\end{center}
Here, $\varepsilon_\F(h) : \Y(h) \circ \F(l_1) \to \F(l_2) \circ
\X(h)$ and $\varepsilon_\G(h) : \Y(h) \circ \G(l_1) \to \G(l_2) \circ
\X(h)$ are the natural isomorphisms witnessing that $\F$ and $\G$ are
themselves degeneracy-preserving.
\end{definition}
\noindent
By contrast with the diagram in Definition~\ref{def:cubical-functor}, the one in
Definition~\ref{def:cubical-nat-trans} is required to commute on the
nose.  

In the setting of \cite{param_johann}, a fibred natural transformation
between two fibred functors $\F,\G : |\rel(U)|^n \to \rel(U)$ induces
a face map-preserving cubical natural transformation from $\F$ to $\G$
presented as cubical functors. We note, however, that there is no
notion in~\cite{param_johann} that induces cubical natural
transformations that are {\em both} face map- {\em and}
degeneracy-preserving; indeed, the framework of~\cite{param_johann}
does not require natural transformations between equality-preserving
fibred functors from $|\rel(U)|^n$ to $\rel(U)$ to themselves be
equality-preserving. Significantly, the results of~\cite{param_johann}
can still be obtained even if fibred natural transformations {\em are}
required to be equality-preserving. The analogous requirement for
cubical natural transformations --- namely, the requirement that
cubical natural transformations be both face map- {\em and}
degeneracy-preserving --- thus generalizes structure already present
in the $1$-dimensional setting of~\cite{param_johann}. We exploit this
requirement on cubical natural transformations to construct $\lambda
2$-fibrations in which terms of System F are interpreted as face map-
{\em and} degeneracy-preserving cubical natural transformations. Even
when $p = 1$ this gives a stronger result than is obtained
in~\cite{param_johann}, where natural transformations interpreting
terms are not shown to be equality-preserving.


\section{Fibrational models of higher-dimensional
  parametricity}\label{sec:model}
	
In this section we assume a fixed ($p$-dimensional) cubical category
$\rel$. To ensure that composition (and thus substitution) in the base
category of the $\lambda 2$-fibrations we construct is well-defined, as
well as that the cubical functors interpreting System F types will be
not just face map-preserving but also degeneracy-preserving, we need
to consider equality of morphisms up to (certain kinds of) natural
isomorphisms.  We therefore assume that our fixed cubical category
$\rel$ comes equipped with a class $M =
{\mathlarger{\mathlarger{\Sigma}}}_{l \leq p} M(l)$ of ``good''
isomorphisms of the form $J \to \rel(l)_1$, where $J$ is some object
of $\C$ and $\rel(l)_1$ is the object of morphisms in the internal
category $\rel(l)$. We require that each $M(l)$ contains all identity
morphisms in $\rel(l)$, is closed under composition and inverses in
$\rel(l)$, and is closed under reindexing by morphisms in $\C$.  The
isomorphisms in $M$ will be used below to parameterize various
constructions on cubical categories over different notions of
equivalence of morphisms. For example, taking $M$ to be the class of
identity morphisms will ensure that diagrams commute on the nose ---
as for Bainbridge {\em et al.}'s PER model; see Example~\ref{ex:per}
below --- while taking $M$ to be the class of all isomorphisms will
entail that the same diagrams commute up to an arbitrary natural
isomorphism. Less extremal choices for $M$ are possible as well: for
example, we can define $M$ by induction on $l$, letting $M(0)$ consist
of only the identity morphisms, and defining an isomorphism $f$ to
belong to $M(l+1)$ iff $\rel(h) \circ f$ belongs to $M(l)$ for any
face map $h : l+1 \to l$, where this composition is in $\C$.  This
definition of $M$ is used for Reynolds' model; see
Example~\ref{ex:reynolds}.

We will use the following cubical categories --- one for each $n$ ---
to interpret System F types:

\begin{definition}\label{def:reln-to-rel}
The ($p$-dimensional) cubical category $|\rel|^n \to \rel$ is
given as follows:
\begin{itemize}
\item the objects are triples $(\F,\varepsilon_\F,\upsilon_\F)$, where  
\begin{itemize}
\item $\F$ is a face map-preserving ($p$-dimensional) cubical functor
  from $|\rel|^n$ to $\rel$ 
\item $\varepsilon_\F$ is a family of natural isomorphisms witnessing
  that $\F$ is also degeneracy-preserving. Moreover,
  $\varepsilon_\F(h)$ is in $M(l_2)$ for each degeneracy $h: l_1 \to
  l_2$ in $\square_p$
\item $\upsilon_\F$ is a function associating to each isomorphism $f :
  \rel(l)^m_0 \to \rel(l)^n_1$ with the property that $\pi_k \circ f$
  is in $M(l)$ for each $k \leq n$ an isomorphism $\upsilon_\F(f) :
  \rel(l)^m_0 \to \rel(l)_1$ in $M(l)$. Moreover, $\upsilon$ respects
  the source and target operations, as well as identities,
  composition, and reindexing of isomorphisms
\end{itemize}
\item the morphisms are face map- and degeneracy-preserving
  ($p$-dimensional) cubical natural transformations 
\end{itemize} 
\end{definition}

Generalizing from the $1$-dimensional setting of~\cite{param_johann},
in which types with $n$ free variables are interpreted as
equality-preserving functors from $|\rel(1)|^n$ to $\rel(1)$, 
we aim to interpret System F types as face map- and
degeneracy-preserving cubical functors from $|\rel|^n$ to $\rel$ for
various $n$; the restriction to functors with discrete domains here
makes it possible to handle all type expressions in System F, not just
the positive ones. This will require that the total categories of the
$\lambda 2$-fibrations we construct have such functors as their
objects. But for such functors to form a category, they must support a
well-defined notion of composition. To ensure that this is the case
even though cubical functors are only required to preserve
degeneracies up to isomorphism, and even though those isomorphisms
are, importantly, in $\rel$ rather than in $|\rel|$, we need to
arrange that cubical functors from $|\rel|^n$ to $\rel$ for various $n$ preserve
enough isomorphisms. The functions $\upsilon_\F$ accomplish just this:
they endow each cubical functor $\F : |\rel|^n \to \rel$ with enough
structure to preserve all ``good'' isomorphisms, and this is what we
need to push all of the constructions we require through. Of course,
if $\F$ were a cubical functor with domain $\rel^n$ rather than
$|\rel|^n$ we would get the preservation of (all) isomorphisms in
$\rel^n$ for for free. However, this would make it impossible to handle
contravariant type expressions.

When giving a categorical interpretation of System F, a category for
interpreting type contexts is required.
We therefore associate a category of contexts to each cubical
category.

\begin{definition}\label{def:contexts}
 The
($p$-dimensional) {\em category of contexts} $\ctx(\rel)$ is given as
follows:
\begin{itemize}
\item the objects are natural numbers 
\item the morphisms from $n$ to $m$ are $m$-tuples of objects in $|\rel|^n \to \rel$
\end{itemize}
\end{definition}

Defining the product $m \times 1$ in $\ctx(\rel)$ to be the natural
number sum $m + 1$, we see that $\ctx(\rel)$ enjoys sufficient
structure to model the construction of System F type contexts:

\begin{lemma}
The category $\ctx(\rel)$ has a terminal object 0 and a choice of
products $(-) \times 1$.  
\end{lemma}


To appropriately interpret arrow types will we need to know that each
cubical category of the form $|\rel|^n \to \rel$ is cartesian
closed. The next three lemmas show that, under reasonable conditions
on $\rel$, this is indeed the case. The constructions are variants of
familiar ones, except that care must be taken to ensure that the
isomorphisms in $M$ are respected.

\begin{definition}\label{def:terminal}
$\rel$ \emph{has terminal objects} if it comes equipped with a choice
  of terminal objects $1_l$ in $\rel(l)$ for $l \leq p$. This choice
  of terminal objects is \emph{stable under face maps} if the equality
  below holds for each face map $h : l_1 \to l_2$ in $\square_p$:
\[\rel(h) \, 1_{l_1} = 1_{l_2}\]
It is \emph{stable under degeneracies} if the equality holds up to an
isomorphism in $M(l_2)$ for each degeneracy $h : l_1 \to l_2$ in
$\square_p$.
\end{definition}

\noindent
We write $1$ rather than $1_l$ below when $l$ is clear from context.

\begin{lemma}\label{lem:terminal-choice}
If $\rel$ has terminal objects that are stable under face maps and
degeneracies then we have a choice of terminal objects $1_n$ in
$|\rel|^n \to \rel$.
\end{lemma}

\begin{definition}\label{def:products}
$\rel$ \emph{has products} if it comes equipped with a choice of
  products $(\times_l,\fst_l,\snd_l)$ in $\rel(l)$ for $l \leq p$,
  such that $M(l)$ for $l \leq p$ is closed under products.  This
  choice of products is \emph{stable under face maps} if, for each
  face map $h : l_1 \to l_2$ in $\square_p$, the equalities below hold
  for any objects $A,B$ of $\rel(l_1)$:
\begin{align*}
& \rel(h) \; (A \times_{l_1} B) = \big(\rel(h) \; A\big) \times_{l_2}
  \big(\rel(h) \; B\big) \\ 
& \rel(h) \; (\fst_{l_1}[A,B]) = \fst_{l_2}\big[\rel(h) \; A, \rel(h) \; B\big] \\
& \rel(h) \; (\snd_{l_1}[A,B]) = \snd_{l_2}\big[\rel(h) \; A, \rel(h) \; B\big]
\end{align*}
It is \emph{stable under degeneracies} if, for each degeneracy $h :
l_1 \to l_2$ in $\square_p$, the first equality above holds up to an
isomorphism $\varepsilon(h,A,B)$ in $M(l_2)$ that makes the following
two diagrams commute:
\begin{center}
\scalebox{0.75}{
\begin{tikzpicture}
\node (N0) at (0,3) {$\rel(h) \; (A \times_{l_1} B)$};
\node (N1) at (3.5,0) {$\rel(h) \; A$};
\node (N2) at (7,3) {$\big(\rel(h) \; A\big) \times_{l_2} \big(\rel(h) \; B\big)$};
\draw[->] (N0) -- node[below]{$\rel(h) \; (\fst_{l_1}[A,B])$ \;\;\;\;\;\;\;\;\;\;\;\;\;\;\;\;\;\;\;\;\;\;\;\;\;} (N1);
\draw[->] (N0) -- node[above]{$\varepsilon(h,A,B)$} (N2);
\draw[->] (N2) -- node[below]{\;\;\;\;\;\;\;\;\;\;\;\;\;\;\;\;\;\;\;\;\;\;\;\;\;\;\;\;\;\; $\fst_{l_2}\big[\rel(h) \; A, \rel(h) \; B\big]$} (N1);
\end{tikzpicture}}
\end{center}

\begin{center}
\scalebox{0.75}{
\begin{tikzpicture}
\node (N0) at (0,3) {$\rel(h) \; (A \times_{l_1} B)$};
\node (N1) at (3.5,0) {$\rel(h) \; B$};
\node (N2) at (7,3) {$\big(\rel(h) \; A\big) \times_{l_2} \big(\rel(h) \; B\big)$};
\draw[->] (N0) -- node[below]{$\rel(h) \; (\snd_{l_1}[A,B])$ \;\;\;\;\;\;\;\;\;\;\;\;\;\;\;\;\;\;\;\;\;\;\;\;\;} (N1);
\draw[->] (N0) -- node[above]{$\varepsilon(h,A,B)$} (N2);
\draw[->] (N2) -- node[below]{\;\;\;\;\;\;\;\;\;\;\;\;\;\;\;\;\;\;\;\;\;\;\;\;\;\;\;\;\;\; $\snd_{l_2}\big[\rel(h) \; A, \rel(h) \; B\big]$} (N1);
\end{tikzpicture}}
\end{center}
\end{definition}

\noindent
We write $(\times,\fst,\snd)$ rather than $(\times_l,\fst_l,\snd_l)$
when $l$ is clear from context.

\begin{lemma}\label{lem:products-choice}
If $\rel$ has products stable under face maps and degeneracies then we
have a choice of products $(\times_n,\fst_n,\snd_n)$ in $|\rel|^n \to
\rel$.
\end{lemma}

If, following the development in~\cite{param_johann}, we did not
require cubical natural transformations to preserve degeneracies, then
we would not need to require commutativity of the two diagrams above
for degeneracies. We would still need Definition~\ref{def:products}'s
requirement on face maps, however.

\begin{definition}\label{def:exponentials}
$\rel$ \emph{has exponentials} if it has products and it comes
  equipped with a choice of exponentials $(\Rightarrow_l,\eval_l)$ in
  $\rel(l)$ for $l \leq p$ with respect to the chosen products, such
  that $M(l)$ for $l \leq p$ is closed under exponentials. This
  choice of exponentials is \emph{stable under face maps} if the
  choice of products is stable under face maps and, for each face map
  $h : l_1 \to l_2$ in $\square_p$, the equalities below hold for any
  objects $A,B$ of $\rel(l_1)$:
\begin{align*}
& \rel(h) \; (A \Rightarrow_{l_1} B) = \big(\rel(h) \; A\big)
  \Rightarrow_{l_2} \big(\rel(h) \; B\big) \\ 
& \rel(h) \; (\eval_{l_1}[A,B]) = \eval_{l_2}\big[\rel(h) \; A,
    \rel(h) \; B\big] 
\end{align*}
It is \emph{stable under degeneracies} if the choice of products is
stable under degeneracies and, for each degeneracy $h : l_1 \to l_2$
in $\square_p$, the first equality above holds up to an isomorphism
$\upsilon(h,A,B)$ in $M(l_2)$ that makes the following diagram
commute:
\begin{center}
\scalebox{0.6}{
\begin{tikzpicture}
\node (N0) at (0,6) {$\rel(h) \; \big((A \Rightarrow_{l_1} B) \times_{l_1} A\big)$};
\node (N1) at (9,0) {$\rel(h) \; B$};
\node (N2) at (9,6) {$\big(\rel(h) \; (A \Rightarrow_{l_1} B)\big) \times_{l_2} \big(\rel(h) \; A\big)$};
\node (N3) at (9,3) {\;\;\;\;\; $\Big(\big(\rel(h) \; A\big) \Rightarrow_{l_2} \big(\rel(h) \; B\big)\Big) \times_{l_2} \big(\rel(h) \; A\big)$};
\draw[->] (N0) -- node[below]{$\rel(h) \; (\eval_{l_1}[A,B])$ \;\;\;\;\;\;\;\;\;\;\;\;\;\;\;\;\;\;\;\;\;\;\;\;\;} (N1);
\draw[->] (N0) -- node[above]{$\varepsilon(h,A \Rightarrow_{l_1} B,A)$} (N2);
\draw[->] (N2) -- node[right]{$\upsilon(h,A,B) \times_{l_2} 1$} (N3);
\draw[->] (N3) -- node[right]{$\eval_{l_2}\big[\rel(h) \; A, \rel(h) \; B\big]$} (N1);
\end{tikzpicture}}
\end{center}
Here, $\varepsilon(h,A \Rightarrow_{l_1} B,A)$ is the isomorphism in $M(l_2)$ witnessing the stability of the product in
question under $h$.
\end{definition}

\noindent
We write $(\Rightarrow, \eval)$ rather than $(\Rightarrow_l, \eval_l)$
when $l$ is clear from context.

\begin{lemma}\label{lem:exponentials-choice}
If $\rel$ has exponentials stable under face maps and degeneracies
then we have a choice of exponentials
$(\Rightarrow_n,$ $\eval_n)$ in $|\rel|^n \to \rel$.
\end{lemma}

As for products, if we follow the development in~\cite{param_johann}
and did not require cubical natural transformations to preserve
degeneracies, then we would not need to require commutativity of the
above diagram for degeneracies. We would still need
Definition~\ref{def:exponentials}'s requirement on face maps, however.

Putting Lemmas~\ref{lem:terminal-choice},~\ref{lem:products-choice},
and~\ref{lem:exponentials-choice} together gives:

\begin{proposition}
If a cubical category $\rel$ has terminal objects, products, and
exponentials, all of which are stable under face maps and
degeneracies, then $|\rel|^n \to \rel$ is cartesian closed.
\end{proposition}

In the development above we consider cubical categories to be functors
with codomain $\cat$, as explained in
Section~\ref{sec:cubical-categories} above. If, however, we more
properly view $p$-dimensional cubical categories as functors from
$\square_p$ to $\cat(\C)$, then the construction of terminal
objects, products, and exponentials must actually be carried out
internally to our ambient category $\C$. This means that in
Definition~\ref{def:products}, for example, $A$ and $B$ are morphisms
into $\C$'s object of objects, and their product is an internal
product. The necessary definition of internal products is standard and
can be found, for example, in Section 7.2 of~\cite{jac99}. A similar
remark applies to Definitions~\ref{def:terminal}
and~\ref{def:exponentials}, and at several places below, but we will
suppress remarks analogous to this one in the remainder of this paper.

\vspace*{0.1in}

The cubical category $|\rel|^n \to \rel$ will ultimately emerge as the
fiber over object $n$ of $\ctx(\rel)$ in the $\lambda 2$-fibration we
construct to interpret System F. To interpret $\forall$-types we will
require a right adjoint to context weakening that moves between such
fibers and is appropriate to the cubical setting. To formalize this
requirement, we first define the category that will be the total
category of our $\lambda 2$-fibration.

\begin{definition}
The ($p$-dimensional) cubical category $\int_n \, |\rel|^n \to \rel$ is
given as follows:
\begin{itemize}
\item the objects are pairs $(n,\F)$, where $\F$ is an object in
  $|\rel|^n \to \rel$
\item the morphisms from $(n,\F)$ to $(m,\G)$ are pairs
  $(\mathbf{F},\eta)$, where $\mathbf{F} : n \to m$ is a morphism in
  $\ctx(\rel)$ and $\eta : \F \to \G \circ \mathbf{F}$ is a morphism
  in $|\rel|^n \to \rel$
\end{itemize} 
\end{definition}

Since the set of objects of $|\rel|^n \to \rel$ is, by definition,
(isomorphic to) the set of morphisms $\Mor(n,1)$ in $\ctx(\rel)$, we
have not only that $\int_n \, |\rel|^n \to \rel$ is the total category
of a fibration over $\ctx(\rel)$, but that this fibration is actually
a split fibration.

\begin{lemma}
The forgetful functor from $\int_n \, |\rel|^n \to \rel$ to
$\ctx(\rel)$ is a split fibration with split generic object $1$.
\end{lemma}

\noindent
Moreover, cartesian structure from $\rel$ lifts to this fibration:

\begin{lemma}\label{lem:split}
If $\rel$ has terminal objects, products, and exponentials, all stable
under face maps and degeneracies, then the forgetful functor from
$\int_n \,|\rel|^n \to \rel$ to $\ctx(\rel)$ is a split cartesian
closed fibration with split generic object $1$.
\end{lemma}

\vspace*{0.05in}

The split cartesian closed structure identified in Lemma~\ref{lem:split}
will allow us to interpret of
function types. To ensure that we can also interpret $\forall$-types
we require some additional structure.

\begin{definition}\label{def:split-simple-products}
Let $U : \mathcal{E} \to \mathcal{B}$ be a split fibration with a
distinguished object $\Omega$ of $\mathcal{B}$ and a choice of
products $(-) \times \Omega$ in $\mathcal{B}$. We say that $U$
\emph{has split simple $\Omega$-products} if it comes equipped with a
choice of right adjoints $\forall_A : \mathcal{E}_{A \times \Omega}
\to \mathcal{E}_A$ to the weakening functors $\fst[A, \Omega]^* :
\mathcal{E}_A \to \mathcal{E}_{A \times \Omega}$ for objects $A$ of
$\mathcal{B}$, with the respective unit and counit pairs
$(\eta_A,\varepsilon_A)$, satisfying the following conditions for
every morphism $f : A \to B$ in $\mathcal{B}$:
\begin{itemize}
\item the following diagram commutes:
\begin{center}
\scalebox{0.75}{
\begin{tikzpicture}
\node (N0) at (0,2) {$\mathcal{E}_{B \times \Omega}$};
\node (N1) at (3,2) {$\mathcal{E}_B$};
\node (N2) at (0,0) {$\mathcal{E}_{A \times \Omega}$};
\node (N3) at (3,0) {$\mathcal{E}_A$};
\draw[->] (N0) -- node[above]{$\forall_B$} (N1);
\draw[->] (N0) -- node[left]{$(f \times 1)^*$} (N2);
\draw[->] (N2) -- node[below]{$\forall_A$} (N3);
\draw[->] (N1) -- node[right]{$f^*$} (N3);
\end{tikzpicture}}
\end{center}
\item $f^*(\eta_B(X)) = \eta_A(f^*(X))$ for every object $X$ of
  $\mathcal{E}(B)$
\item $(f \times 1)^*(\varepsilon_B(X)) = \varepsilon_A((f \times
  1)^*(X))$ for every object $X$ of $\mathcal{E}_{B \times \Omega}$
\end{itemize}
\end{definition}


Fibrations with enough structure to give sound interpretations of
System F were dubbed ``$\lambda 2$-fibrations'' by
Seely~\cite{see87}:

\begin{definition}
A \emph{split $\lambda 2$-fibration} is a split cartesian closed
fibration $U : \mathcal{E} \to \mathcal{B}$, that has a terminal
object in $\mathcal{B}$, a split generic object $\Omega$, chosen
products $(-)\times \Omega$ in $\mathcal{B}$, and split simple
$\Omega$-products.
\end{definition}

\begin{definition}\label{def:parametric-model}
$\rel$ is a {\em ($p$-dimensional) parametric model} of System F if it
  has terminal objects, products, and exponentials, all stable under
  face maps and degeneracies, and is such that the forgetful functor
  from $\int_n \,|\rel|^n \to \rel$ to $\ctx(\rel)$ has split simple
  $1$-products.
\end{definition}

Our main technical theorem shows that every parametric model of System
F naturally gives rise to a split $\lambda 2$-fibration. The
construction is also a careful variant of familiar ones.

\begin{theorem}\label{thm:split-lambda2-fib}
If $\rel$ is a ($p$-dimensional) parametric model of System F, then
the forgetful functor from $\int_n \,|\rel|^n \to \rel$ to
$\ctx(\rel)$ is a split $\lambda 2$-fibration.
\end{theorem}

We also have the following variant of Seely's~\cite{see87} result that
every split $\lambda2$-fibration gives rise to a sound model of System
F:

\begin{theorem}\label{prop:model}
Every split $\lambda 2$-fibration $U : \mathcal{E} \to \mathcal{B}$
gives a model of System F in which:
\begin{itemize}
\item every type context $\Gamma$ is interpreted as an object
  $\sem{\Gamma}$ in $\mathcal{B}$
\item every type $\Gamma \vdash T$ is interpreted as an object
  $\sem{\Gamma \vdash T}$ in the fiber $\mathcal{E}_{\sem{\Gamma}}$
\item every term context $\Gamma;\Delta$ is interpreted as an object
  $\sem{\Gamma \vdash \Delta}$ in the fiber $\mathcal{E}_{\sem{\Gamma}}$
\item every term $\Gamma;\Delta \vdash t : T$ is interpreted as a
  morphism $\sem{\Gamma;\Delta \vdash t : T}$ from
  $\sem{\Gamma;\Delta}$ to $\sem{\Gamma \vdash T}$ in the fiber
  $\mathcal{E}_{\sem{\Gamma}}$
\end{itemize}
Moreover, if $\Gamma;\Delta \vdash s =_{\beta\eta} t : T$, then
$\sem{\Gamma;\Delta \vdash s : T} = \sem{\Gamma;\Delta \vdash t : T}$.
\end{theorem}

Theorems~\ref{thm:split-lambda2-fib} and~\ref{prop:model} together
imply our main result, namely:

\begin{theorem}\label{thm:cubical-model}
A ($p$-dimensional) parametric model $\rel$ of System F gives a sound
model of System F in which
\begin{itemize}
\item every type $\Gamma \vdash T$ is interpreted as a face map- and
  degeneracy-preserving cubical functor $\sem{\Gamma \vdash T} :
  |\rel|^{|\Gamma|} \to \rel$
\item every term $\Gamma; \Delta \vdash t : T$ is interpreted as a
  face map- and degeneracy-preserving cubical natural transformation
  $\sem{\Gamma; \Delta \vdash t:T} : \sem{\Gamma \vdash \Delta} \to
  \sem{\Gamma \vdash T}$
\end{itemize}
\end{theorem}

Taking $p = 1$ and omitting the requirement that cubical natural
transformations be degeneracy-preserving as indicated at several
places above shows that Theorem~\ref{thm:cubical-model} naturally
generalizes Theorem~4.6 of~\cite{param_johann} to arbitrary (including
infinite, when $p = \infty$) higher dimensions. In particular, the
fact that our cubical functors interpreting types are
degeneracy-preserving gives a higher-dimensional analogue of the
fibrational formulation of Reynolds' Identity Extension Lemma
from~\cite{param_johann}.

\section{Examples}\label{sec:examples}

In this section we show how both Reynolds' original model and the PER
model of Bainbridge {\em et al.} arise as instances of our theory.

\begin{example}\label{ex:reynolds}
We consider Reynolds' original model, which is internal to the
Calculus of Inductive Constructions with Impredicative $\set$. In the
interest of clarity, we write $\U$ (rather than $\set$, as in
implementations of Coq) for the impredicative universe $\U$. We then
define
\begin{align*}
& \mathsf{isProp}(A) \coloneqq \Pi_{a,b:A} \, \Id(a,b) \\
& \Prop \coloneqq \Sigma_{A : \U} \, \mathsf{isProp}(A) \\
& \mathsf{isSet}(A) \coloneqq \Pi_{a,b:A} \, \mathsf{isProp}(\Id(a,b)) \\
& \Set \coloneqq \Sigma_{A : \U} \, \mathsf{isSet}(A)
\end{align*} 
Here, $\Sigma$ forms dependent sums, $\Pi$ forms dependent products,
and $\Id$ is the identity type.  Intuitively, $\set$ is the type of
types in $\U$ that are ``discrete''.  We therefore treat the terms of
$\Set$ as if they were types in $\U$.  Since $\U$ is impredicative, we
have $\Set : \U$.

To capture Reynolds' construction we take our ambient category $\C$ to
be the category whose objects are the types in $\U$, and whose
morphisms are equivalence classes of functions. Here, functions $f,g :
A \to B$ are considered equal precisely when the type $\eq(f,g)$ is
inhabited. To keep from incorporating any particular computational
structure into the categorical structure, it is crucial that we use
proof-irrelevant propositional equality types $\mathsf{eq}(-,-)$,
rather than proof-relevant identity types $\Id(-,-)$, here; this
ensures, for example, that the uniqueness condition for pullbacks is
satisfied. With this definition it is easy to check that $\C$ is
finitely complete.

To see the type $\Set$ as a category internal to $\C$ we first
define the type $\Set(A,B)$ of morphisms from $A$ to $B$ to be
$\Set(A,B) \coloneqq A \to B$, and then take the object of objects in
the internal category to be $\Set$ itself and its object of morphisms
to be $\Sigma_{A, B: \U} A \to B$. We define a category of relations
by
\begin{align*}
& \mathsf{R} \coloneqq \Sigma_{A,B: \Set} A \times B \to \Prop \\
& \mathsf{R} \; (A_1,A_2,R_A) \; (B_1,B_2,R_B) \coloneqq \Sigma_{f :
    A_1 \to B_1} \Sigma_{g : A_2 \to B_2} \\ 
& \;\;\;  \Pi_{a_1:A_1} \Pi_{a_2 : A_2} R_A(a_1,a_2) \to R_B(f a_1 ,g a_2)
\end{align*}
which we can see as a category internal to $\C$ whose object of
objects is $\mathsf{R}$ itself and whose object of morphisms is
\begin{align*}
& \Sigma_{(A_1,A_2,R_A),(B_1,B_2,R_B) :\mathsf{R}}\, \Sigma_{f :
    A_1 \to B_1} \Sigma_{g : A_2 \to B_2}\\ 
& \;\;\;\;\;\;\;\;\; \Pi_{a_1:A_1} \Pi_{a_2 : A_2} R_A(a_1,a_2) \to R_B(f
  a_1 ,g a_2) 
\end{align*}
We obviously have two internal functors from $\mathsf{R}$ to $\Set$
corresponding to the first and second projections, respectively. We
also have an equality functor $\Eq$ from $\Set$ to $\mathsf{R}$
defined by 
\begin{align*}
& \Eq \,A \coloneqq (A,A,\Id_A) \\
& \Eq \, f \coloneqq (f,f,\mathsf{ap}\,f)
\end{align*}
where $\mathsf{ap}\,f : \Id_A (a_1,a_2) \to \Id_B(f a_1, f a_2)$ is
defined as usual by $\Id$-induction.

We obtain a $1$-dimensional cubical category $\rel$ by defining
$\rel(0) = \set$ and $\rel(1) = \mathsf{R}$, and mapping the two face
maps to the two projections, and the single degeneracy to $\Eq$. We
can define terminal objects, products, and exponentials for $\rel$ in
the obvious ways, relating two pairs iff their first and second
components are related, and two functions iff they map related
arguments to related values. It is not hard to check that all these
constructs are preserved on the nose by the two face maps
(projections), and preserved up to a natural isomorphism whose first
and second projections are identities by the single degeneracy
(equality functor). All three constructs are therefore stable under
both face maps and degeneracies. As noted in the introduction, the
difference between fibred functors preserving equality on the nose or
only up to natural isomorphism is precisely where the construction
in~\cite{param_johann} fails. Composition and substitution in (what is
intended to be) the base category of the $\lambda 2$-fibration
constructed in the main theorem there cannot be defined in any
standard way unless equality is preserved on the nose, but equality
in~\cite{param_johann} is only defined --- and therefore can only be
preserved --- up to isomorphism.

Finally, we define the adjoint $\forall_n$ by 
\begin{align*}
&\forall_n \,\F(0) \; \overline{A} \coloneqq \Sigma_{f : \Pi_{A:\Set}
    \F(0)(\overline{A},A)} \Pi_{R : \mathsf{R}} \\ 
& \;\;\;\;\; \pi_3\,\big(\F(1)(\overline{\Eq \,A},R) \; (f \;
  (\pi_1R))' \; (f \; (\pi_2R))'\big) 
\end{align*}
\begin{align*}
&\forall_n \,\F(1) \; \overline{R} \coloneqq \Big(\forall_n\,\F(0)
  \; \overline{\pi_1(R)}, \forall_n\,\F(0) \; \overline{\pi_2(R)}, \\
& \;\;\; \lambda_{f : \forall_n \F(0) \; \overline{\pi_1 \,R}} \; \lambda_{g : \forall_n \F(0) \;
    \overline{\pi_2 R}} \\ 
& \;\;\;\;\; \pi_3(\F(1)(R,R)) \; (f \; (\pi_1 R))'
  \; (g \; (\pi_2 R))'\Big) 
\end{align*}
In the above, the term $(f \, (\pi_1R))' :
\pi_1\big(\F(1)(\overline{\Eq \,A},R\big))$ stands for the term $f
\, (\pi_1R) : \F(0)(\overline{A},\pi_1R))$ transported along the
equality between the respective types, and similarly for $\pi_2$ and
$g$. We emphasize again that these terms all exist because $\rel$
preserves face maps on the nose.
\end{example}

\begin{example}\label{ex:per}
We consider the PER model of Bainbridge {\em et al.} internal
to the category of $\omega$-sets. We follow the development
of~\cite{longo_moggi} for concepts related to this category, In
particular, this category is defined in Definition~6.3
of~\cite{longo_moggi}, and proved in Corollary~8.3 there to be
finitely complete.

We construct a $1$-dimensional cubical functor $\rel$ as follows. As
our internal category $\rel(0)$ of $0$-relations we take the category
$\mathbf{M}'$ as in Definition 8.4 of~\cite{longo_moggi}. Informally,
the objects $\mathbf{M}'$ are partial equivalence relations, and its
morphisms are realizable functions that respect those relations.  

To define the internal category $\rel(1)$ of $1$-relations, we first
construct its object of objects.  As the carrier of this $\omega$-set
we take the set of triples $(A, B, R)$, where $A$ and $B$ are partial
equivalence relations and $R$ is a saturated predicate on $A \times
B$. Here the product $A \times B$ of two PERs is constructed in the
standard way, using a bijective pairing function $\langle
\cdot,\cdot\rangle$ and relating two pairs iff their respective
projections --- which we will call $\fst$ and $\snd$ below --- are
related. A saturated predicate on a PER $A$ is a predicate $R$ on
natural numbers that is closed under $A$, in the sense that $m \sim_A
n$ and $R(m)$ imply $R(n)$. To finish the construction of our object
of objects for $\rel(1)$ we take any triple $(A, B, R)$ as above to be
realized by any natural number.

As the carrier of the object of morphisms for $\rel(1)$ we take the
set of quadruples of the form \[\big((A_1,B_1,R_1), (A_2,B_2,R_2),
\{n\}_{A_1 \to A_2}, \{m\}_{B_1 \to B_2}\big)\] satisfying the
condition that, for any $k$ such that $R_1(k)$ holds, we have that
$R_2\big(\langle n \cdot \fst(k), m \cdot \snd(k) \rangle\big)$ holds
as well.  The first two components of such a quadruple serve to encode
the domain and codomain of the morphism. The third component is a
(nonempty) equivalence class under the exponential PER $A_1 \to
A_2$. Here the exponential $A \to B$ of two PERs is constructed in the
standard way, using an encoding of partial recursive functions as
natural numbers and relating two functions iff they map related
arguments to related values. In accordance with \cite{longo_moggi}, we
denote the application of the $n^{th}$ partial recursive function to a
natural number $a$ in its domain by $n \cdot a$. To finish the
construction of our object of morphisms for $\rel(1)$, we take a
quadruple as above to be realized by a natural number $k$ iff $\fst(k)
\sim_{A_1 \to A_2} n$ and $\snd(k) \sim_{B_1 \to B_2} m$.

We obviously have two internal functors from $\rel(1)$ to $\rel(0)$,
corresponding to the first and second projections, respectively. We
also have an equality functor $\mathsf{Eq}$ from $\rel(0)$ to
$\rel(1)$ whose action on objects is given by $\mathsf{Eq}\,A
\coloneqq (A,A,R_A)$, where $R_A(k)$ iff $\fst(k) \sim_A \snd(k)$, and
whose action on morphisms is given by $\mathsf{Eq}\,(A,B,\{n\}_{A \to
  B}) \coloneqq (\Eq\,A,\Eq\,B, \{n\}_{A \to B},\{n\}_{A \to B})$. We
therefore have that $\rel$ is indeed a $1$-dimensional cubical
category. We can define terminal objects, products, and exponentials
for $\rel$ in the obvious ways, inheriting from the corresponding
standard constructs on PERs. It is not hard to check that all these
constructs are preserved both by the two face maps (projections), and
by the single degeneracy (equality functor), on the nose.

Finally, we define the adjoint $\forall_n$ on objects by
\begin{align*}
&\forall_n\,\F(0) \; \overline{A} \coloneqq \big\{ (n,k) \; | \;
  \text{ for all} \; A : \mathbf{M}', n \sim_A k, \\ & \;\;\;
  \text{and for all} \; R : \rel(1),
  \pi_3\,\big(\F(1)(\overline{\Eq\,A},R)\big) \; (n,k) \big\}
\end{align*}
\begin{align*}
&\forall_n\,\F(1) \; \overline{R} \coloneqq \big(\forall_n\,\F(0)
  \; \overline{\pi_1(R)}, \forall_n\,\F(0) \; \overline{\pi_2(R)}, \\  
& \;\;\; \big\{ n \; | \; \text{ for all} \; R : \rel(1),
  \pi_3\,(\F(1) (\overline{R},R)) \; n \big\} \big)
\end{align*}
\noindent
To define $\forall_n$ on a morphism $\eta : \F \to \G$, we define
\begin{align*}
& \forall_n\,\eta(0) \; \overline{A} \coloneqq \big( \forall_n\,\F(0)
  \; \overline{A}, \forall_n\,\G(0) \; \overline{A}, \\ &
  \;\;\;\;\;\;\;\;\;\;\;\;\;\;\;\;\;\;\;\;\;\;\;\;\; \{ m \cdot 0
  \}_{\forall_n\,\F(0) \; \overline{A} \to \forall_n\,\G(0) \;
    \overline{A}}\big)
\end{align*}
\noindent
Here, $m$ is any natural number realizing $\eta(0) \bar{A}$.  It is
crucial that all natural transformations are ``uniformly realized", in
the sense that there is a natural number realizing each such
transformation and, because all PERs are defined to be realized by all
natural numbers, each is suitably uniform. In particular, if $\eta$
were not uniformly realized in the above sense, then $\forall$ would
not be well-defined. Using this observation it is possible to show
that, in the category-theoretic setting (rather than in the setting of
$\omega$-sets), the adjoint $\forall_n$ cannot exist precisely because
{\em ad hoc} natural transformations --- {\em i.e.}, natural
transformations that are not uniformly realizable, even though each of
their components may indeed be realizable --- are not excluded.
\end{example}

\section{Consequences of Parametricity}\label{sec:consequences}

In this section we show that the models constructed in
Theorem~\ref{thm:cubical-model} satisfy the properties that ``good''
models of parametricity for System F should satisfy. In particular,
Lemma~\ref{lem:graph-lemma} below shows that, under reasonable
conditions, our models support the definition of a graph for each face
map- and degeneracy-preserving cubical functor. Moreover,
Theorem~\ref{thm:initial-algebras-exist} and its analogue for final
coalgebras show that our higher-dimensional models of relational
parametricity for System F also validate the existence of initial
algebras and final coalgebras for such functors. These results serve
as a sanity check for our theory, and show that it is powerful enough
to show that ``good'' models of relational parametricity for System F
can be constructed even at higher dimensions.

\subsection{A Higher-Dimensional Graph Lemma}

Every function $f : A \to B$ between sets $A$ and $B$ defines a graph
relation $\grph{f} = \{(a,b) \,|\, f\, a = b\}$. This observation can
be phrased fibrationally by letting $U : \rel \to \set \times \set$ be
the standard relations fibration on $\set$, and noting that $\grph{f}$
can be obtained by reindexing the equality relation $\Eq \,B$ on
$B$. In~\cite{param_johann}, the notion of a graph was extended to
more general relations fibrations and a Graph Lemma was proved for
their associated models of $1$-dimensional parametricity. In this
subsection we give a natural generalization of the definition of a
graph from~\cite{param_johann} to the higher-dimensional setting, and
prove a Graph Lemma appropriate to this setting. We begin by
introducing the (new) notion of a cubical (bi)fibration.

\begin{definition}
A ($p$-dimensional) cubical category $\rel$ that has products is a
($p$-dimensional) {\em cubical (bi)fibration} if, for each $l < p$,
each functor 
\[\begin{array}{lll}
\fm(l,k) & = & \langle \rel \, (\fm_\bot (l,k)), \rel \, (\fm_\top (l,k))
\rangle\\ 
       & : & \rel(l+1) \to \rel(l) \times \rel(l)
\end{array}\] 
for $k \leq l$ is a (bi)fibration.
\end{definition}

\noindent
As already noted in Example~\ref{ex:bifib-param}, the (bi)fibrations
$\fm(l,k)$ play the role of the relations fibrations
in~\cite{param_johann}, while the $\dm(l,k)$ play the role of equality
functors. When $\rel$ is a cubical (bi)fibration, we have that $\rel
\, (\dm(l,k)) \, A$ is indeed over $(A,A)$ with respect to $\fm(l,k)$
for every object $A$ in $\rel(l)$, and similarly for every morphism in
$\rel(l)$. 

If $\C$ is a category, write $\C^\to$ for the {\em arrow category} of
$\C$, {\em i.e.}, for the category whose objects are morphisms in $\C$
and whose morphisms from $f : A \to B$ to $f' : A' \to B'$ in $\C^\to$
are pairs of morphisms $g : A \to A'$ and $h : B \to B'$ such that $f'
\circ g = h \circ f$. We define the {\em graph functor} for $\rel$ to
be the set of functors $\{\grph{-}_{l,k} \, |\, l < p, \, k \leq l\}$,
where each $\grph{-}_{l,k}$ is defined as follows:

\begin{definition}
Let $\rel$ be a ($p$-dimensional) cubical fibration that has terminal
objects. For every $l < p$ and $k \leq l$, the functor $\grph{-}_{l,k}
: \rel(l)^\to \to \rel(l+1)$ is defined by:
\begin{itemize}
\item if $h : A \to B$ is an object in $\rel(l)^\to$, then
  $\grph{h}_{l,k} = (h,id_B)^* (\rel\, \dm(l,k)\, B)$
\item if $f : A \to B$, $f' : A' \to B'$, and $(g,h) : f \to f'$ is a
  morphism in $\rel(l)^\to$, then $\grph{g,h}_{l,k}$ is the unique
  morphism from $\grph{f}_{l,k}$ to $\grph{f'}_{l,k}$ obtained from
  $(\rel\, \dm(l,k)\,h) \circ \cart{(f,id_B)}$ via
  $\cart{(f',id_{B'})}$
\end{itemize}
\end{definition}
\[
\xymatrix{
 \grph{f}_{l,k} \ar[r]^-{(f,id_B)^{\S}} \ar@{-->}[d]_-{\exists\, !
   \grph{g,h}_{l,k}} & \rel\,\dm(l,k)\,B \ar[d]^-{\rel\,\dm(l,k)\, h}
 \\ 
 \grph{f'}_{l,k}  \ar[r]_-{(f', id_{B'})^{\S}}	& \rel\,\dm(l,k)\, B' \\
}
\]

Intuitively, one of $\fm_\bot (l,k)$ and $\fm_\top (l,k)$ acts as the
$x$-axis, and the other acts as a $y$-axis, for $l$-dimensional graphs
projected onto dimension $k$. Since reindexing preserves identities,
we have that $\grph{id_A}_{l,k} = (id_A,id_A)^*(\rel\,\dm(l,k)\,B) =
\rel\,\dm(l,k)\,B$. This generalizes the observation that $\grph{id_A}
= \Eq\,A$ in the $1$-dimensional setting of~\cite{param_johann}.

We also have the following alternative characterization of the graph
functor when $\rel$ is a bifibration:

\begin{lemma}\label{lem:opfib-char}
If $\rel$ is a ($p$-dimensional) cubical bifibration that has terminal
objects, and if $f : A \to B$, then $\grph{f}_{l,k} =
\Sigma_{(id_A,f)} \, \dm(l,k) \, A$.
\end{lemma}

\noindent
By contrast with the analogous characterization in Lemma 5.2
of~\cite{param_johann}, no Beck-Chevalley condition is required since
the bifibrations $\fm(l,k)$ are postulated here, rather than derived
from more primitive bifibrations as is done there.

We have the following analogue of Lemma~5.3 of~\cite{param_johann}: 

\begin{lemma}
$\grph{-}_{l,k}$ is full and faithful if $\rel \, \dm(l,k)$ is.
\end{lemma}

Together, the (fibrational) definition of the graph functor and its
opfibrational characterization from Lemma~\ref{lem:opfib-char} give
the following Graph Lemma for our higher-dimensional setting:

\begin{lemma}\label{lem:graph-lemma}
(Graph Lemma) Let $\rel$ be a ($p$-dimensional) cubical bifibration
  that has terminal objects and ${\cal F} : \rel \to \rel$ be a
  ($p$-dimensional) face map- and degeneracy-preserving cubical
  functor. For any $l < p$, $f : A \to B$ in $\rel(l)$, and $k \leq
  l$, there exist morphisms
\[\phi_f : \grph{\F(l) f}_{l,k} \to \F(l+1)\grph{f}_{l,k}\]
and
\[\psi_f : \F(l+1)\grph{f}_{l,k} \to \grph{\F(l) f}_{l,k}\]
in $\rel(l+1)$ that are vertical with respect to $\fm(l,k)$.
\end{lemma}


\subsection{Existence of Initial Algebras and Final Coalgebras}

In this subsection we use our Graph Lemma to show that the models
constructed in Theorem~\ref{thm:cubical-model} validate the existence
of initial algebras and final coalgebras for face map- and
degeneracy-preserving cubical functors, and thus for all
interpretations of positive type expressions in System F. Our
constructions naturally extend those in~\cite{param_johann} to the
higher-dimensional setting.

If $\rel$ is a ($p$-dimensional) cubical category and $\F: \rel \to
\rel$ is ($p$-dimensional) cubical functor, then an
\emph{$\F$-algebra} $(A,k_A)$ is a set of pairs $\{(A_l,k_{A_l}) \, |
\, l < p\}$ in which each $A_l$ is an object of $\rel(l)$ and each
$k_{A_l} : \F(l)A_l \to A_l$ is a morphism in $\rel(l)$.  We call the
set $A = \{A_l \, | \, l < p\}$ the \emph{carrier} of the $\F$-algebra
and the set $k_A = \{k_{A_l} \, | \, l < p\}$ its \emph{structure
  map}. A set of morphisms $f = \{f_l:A_l\rightarrow B_l \, | \, l <
p\}$ with each $f_l$ in $\rel(l)$ is an \emph{$\F$-algebra morphism}
$f : (A,k_A) \to (B,k_B)$ if, for each $f_l$ in $f$, $k_{B_l} \circ
(\F(l)f_l) = f_l \circ k_{A_l}$.  An $\F$-algebra $(Z,\mathit{in})$ is
\emph{weakly initial} if, for any $\F$-algebra $(A,k_{A})$, there
exists a mediating $\F$-algebra morphism $\mathsf{fold}\,[A, k_A] :
(Z, \mathit{in}) \rightarrow (A, k_A)$. It is an \emph{initial
  $F$-algebra} if $\mathsf{fold}\,[A, k_A]$ is unique up to
isomorphism. 

Now, every $\lambda 2$-fibration has an associated internal language.
For the $\lambda 2$-fibration we construct in
Theorem~\ref{thm:split-lambda2-fib}, this is a polymorphic lambda
calculus for which each type $\Gamma \vdash \internal{A}$ is given by
a face map- and degeneracy-preserving cubical functor from
$|\rel|^{|\Gamma|}$ to $\rel$, and each term $\Gamma;\Delta \vdash
\internal{t} : A$ is a face map- and degeneracy-preserving cubical
natural transformation between such functors. We can use this internal
language to reason about our models using System F.

Let $\F : \rel \to \rel$ be a ($p$-dimensional) face map- and
degeneracy-preserving cubical functor. A {\em strength} for $\F$ is a 
set $\sigma = \{\sigma_l\, | \, l < p\}$ of families of morphisms
$(\sigma_l)_{A, B} : \expo{A}{B} \to \expo{\F(l)A}{\F(l)B}$ such that
the mapping of cubical functors to their strengths preserves
identities and composition, and, for each $l < p$ and $k \leq l$,
$\fm(l,k)\,(\sigma_{l+1})_{C,D} = ((\sigma_l)_{A,B},
(\sigma_l)_{A',B'})$ if $\fm(l,k)C = (A,B)$ and $\fm(l,k)D =
(A',B')$. A cubical functor with a strength is said to be {\em
  strong}. Because of the discrete domains, $\sigma$ is a cubical
natural transformation from $\expo{\_}{\_}$ to $\expo{\F\_}{\F\_}$ in
$|\rel|^2 \to \rel$. The term $A, B; \cdot \vdash \internal{\sigma} :
(A \to B) \to (\internal{\F}[A] \to \internal{\F}[B])$ represents the
action of $\F$ on morphisms in the internal language.

To see that every face map- and degeneracy-preserving cubical functor
$\F$ has an initial $\F$-algebra we define $Z = \sem{\forall
  X.(\internal{\F}X\rightarrow X)\rightarrow X}$, $\mathit{fold} =
\Lambda A.\, \lambda k : \internal{\F}A \to A.\, \lambda z:Z.\,
z\,A\,k$, $\mathsf{fold}\,[A,k] =
\sem{\mathit{fold}\,\internal{A}\,\internal{k}}$, where $\internal{A}$
and $\internal{k}$ are the internal expressions corresponding to the
components of another $\F$-algebra $(A,k)$, and $\mathit{in} =
\sem{\lambda x.\,\Lambda X.\,\lambda k:\internal{\F}A \to A.\,k\,
  (\internal{\sigma}\,(\mathit{fold} \,X \,k) \,x)}$. Our Graph Lemma
can then be used to extend the $1$-dimensional construction
from~\cite{param_johann} to the higher-dimensional setting:

\begin{theorem}\label{thm:initial-algebras-exist}
If $\rel$ is a ($p$-dimensional) bifibration that has terminal
objects, if $\F : \rel \to \rel$ is a ($p$-dimensional) face map- and
degeneracy-preserving cubical functor, if $\dm(l,k)$ is full for every
$l < p$ and $k \leq l$, and if, for every $l < p$, $\rel(l)$ is
well-pointed, then $(Z,\mathit{in})$ is an initial $\F$-algebra.
\end{theorem}

\noindent
We obtain the analogous result for final $\F$-coalgebras as well. 

\section{Related Work}

The study of parametricity runs both wide and deep. Here, we draw
connections with some of the work most closely related to ours.

Ma and Reynolds~\cite{mr92} gave the first categorical formulation of
relational parametricity. Generalizing from the evident reflexive
graph structure in well-behaved relational models of the simply typed
lambda calculus, they reformulated Reynolds' original notion of
relational parametricity for System F in terms of reflexive graphs of
Seely's PL categories~\cite{see87}; these have sufficient structure to
model the type-dependent aspects of System F as
well. Jacobs~\cite{jac99} later generalized this reformulation,
recasting it in terms of $\lambda 2$-fibrations and parameterizing it
over a ``logic of types'' for the polymorphic type theory. His
Definition~8.6.2 gives an notion of $1$-dimensional relational
parametricity that is ``external'', in the sense that it describes
when an arbitrary $\lambda 2$-fibration carries enough structure to
formalize that some of the specific models he constructs are
``intuitively parametric''. This contrasts with our ``internal''
approach, which starts with some
suitably-structured-but-otherwise-arbitrary components and uses a
particular construction to weave them into $\lambda 2$-fibrations that
are ``intuitively parametric'' in the same sense as Jacobs' models,
except that our models satisfy this property at higher dimensions,
too. Overall, our work can be seen as a first extension to higher
dimensions of a formalism capturing the observation that ``intuitively
parametric'' $\lambda 2$-fibrations
are all generated in essentially the same way.

Ma and Reynolds~\cite{mr92} neither provide models that are
relationally parametric in the sense they define, nor give any
indication how hard such models might be to construct. This led
Robinson and Rosolini~\cite{rr94} to reconsidered Ma and Reynolds'
reformulation of Reynolds' relational parametricity from the point of
view of internal categories. This supports a narrowing of Ma and
Reynolds' framework that is more promising for model
construction. Robinson and Rosolini also use internal categories to
clarify the constructions of~\cite{mr92}; our use of internal
categories to clarify the constructions of~\cite{param_johann} when $p
= 1$ is in the same spirit.

Dunphy and Reddy~\cite{dr04} do not work with internal categories, but
they do use reflexive graphs to model relations and functors between
reflexive graph categories to model types. The framework they develop
is mathematically elegant and powerful enough to derive some expected
consequences of relational parametricity, including the existence of
initial algebras for strictly positive System F type expressions. The
framework of~\cite{param_johann} offers an alternative categorical
approach to relational parametricity formulated in terms of
bifibrations rather than reflexive graphs. It gives a functorial
semantics for System F that derives all of the expected consequences
of parametricity that Birkedal and M{\o}gelberg prove using
Abadi-Plotkin logic~\cite{bm05}, including the existence of initial
algebras for all positive type expressions, rather than just strictly
positive ones. However, the bifibrational framework suffers from the
shortcomings already discussed in this paper.

Cubical sets were originally introduced in the context of algebraic
topology, but have more recently been shown to model homotopy
  type theory~\cite{cubical_coquand,cubical_harper}, an extension of
Martin-L\"of type theory.  A key feature of homotopy type theory is
that functions are infinitely parametric with respect to
propositional equality in a non-trivial way. It is still not fully
established whether this theory supports a well-defined notion of
computation, even for base types such as natural numbers. That it does
is Voevodsky's \emph{homotopy-canonicity conjecture}.

We are not the first consider parametricity at higher dimensions.
In~\cite{param_ghani}, the bifibrational approach to relational
parametricity developed in~\cite{param_johann} was extended to
proof-relevant relations.  This was achieved by extending the
uniformity condition characterizing parametric functions to proofs by
adding a second ``dimension'' of parametricity on top of Reynolds'
standard one that forces the standard uniformity condition to itself
be uniform, in effect requiring that polymorphic programs can be
proved to map related arguments to related results via related
proofs. The resulting construction
delivers a 2-dimensional parametricity theorem appropriate to the
proof-relevant setting. We conjecture that this construction can be
made an instance of our theory. Note, however, that Definition~22
of~\cite{param_ghani} actually needs our more general theory in
which equality can be required to be preserved only up to natural
isomorphism.

\section{Conclusion and Future Work}

In this paper we developed a theory of higher-dimensional relational
parametricity for System F that not only clarifies and strengthens the
results of~\cite{param_johann} when $p = 1$, but also naturally
generalizes Reynolds' original notion of relational parametricity for
System F to higher dimensions. We have also shown that our theory
properly subsumes Reynolds' original model and the PER model of
Bainbridge {\em et al.} as proper instances of our theory when $p=1$,
and that it formalizes notions of {\em proof-relevant parametricity}
(when $p = 2$) and {\em infinite-dimensional parametricity} when ($p =
\infty$) as well. Finally, we have proved that our theory is ``good''
in the sense that it derives higher-dimensional analogues of expected
results for parametric models. In future work we hope to settle our
conjecture that our $\lambda 2$-fibrations are relationally parametric
in the sense of Jacobs' ``external'' notion when $p = 1$, as well as
to generalize this ``external'' notion to relational parametricity
to infinitely many dimensions. We also plan to investigate how our
theory can be instantiated to give new parametric models for System F
at dimension $1$. Finally, we plan to investigate connections between
our theory and proof-relevant parametricity at dimension $2$, and
between our theory and the homotopy-canonicity conjecture when
$p=\infty$.

\vspace*{0.1in}

\noindent
{\bf Acknowledgments} This research is supported, in part, by NSF
awards 1420175 (PJ) and 1545197 (KS).

\vspace*{0.1in}

\bibliographystyle{plain}
\bibliography{references}

\end{document}